\documentclass[aps,prl,twocolumn,superscriptaddress,showpacs,floatfix]{revtex4}
\usepackage{graphicx,amsmath,amssymb,xspace,epsfig,float,multirow}

\begin{document}

\title{Universal contact for a Tonks-Girardeau gas at finite temperature}

\author{Patrizia Vignolo}
\affiliation{Universit\'e de Nice-Sophia
  Antipolis, Institut Non Lin\'eaire de Nice, CNRS, 1361 route des Lucioles, 06560 Valbonne, France}
\author{Anna Minguzzi}
\affiliation{Universit\'e Grenoble I and CNRS, Laboratoire de Physique et
  Mod\'elisation, des Milieux Condens\'es,  UMR 5493, B.P.\ 166, 38042 Grenoble,
  France}

\date{\today}

\begin{abstract}
We determine the finite-temperature momentum distribution of a strongly interacting 1D Bose gas in the Tonks-Girardeau (impenetrable-boson) limit under harmonic confinement, and explore its universal properties associated to the scale invariance of the model. We show that, at difference from the unitary Fermi gas in three dimensions, the weight of its large-momentum tails -- given by the Tan's contact --  increase with temperature, and calculate the high-temperature universal second contact coefficient using a virial expansion.
\end{abstract}


\maketitle

{\it Introduction.} 
Ultracold atomic gases are a unique system to probe universality aspects in physics. Due to their diluteness condition, the interactions are in most of the cases modeled by a single parameter, the $s$-wave scattering length, which describes the only type of collisions allowed at the ultralow temperatures typical of the experimental conditions. Another remarkable aspect is that the interactions are tunable (eg by the mechanism of Feschbach resonances). When the interactions are brought to extremely large -- positive or negative -- values, provided that the system remains stable against collapse or three-body recombination, there is no energy or length scale associated to the interactions. In this regime, characterized by spatial scale invariance, all the properties of the system are governed only by external parameters, such as the temperature, the density or the external confinement,  and the system thus displays {\em universal} features. For the case of a {\em three-dimensional} two-component Fermi gas in the so-called unitary regime (corresponding to infinitely large scattering length)  several universal aspects have been explored both theoretically \cite{Ho2004,Viverit2004,Tan2005,Tan2008a,Tan2008b,Tan2008c,Braaten2008,Werner2009,Zhang2009} and experimentally \cite{OHara2002,Bourdel2003,Kinast2005,Stewart2010,Ku12,VanHou12}. 

We consider here  a {\em one-dimensional} (1D) Bose gas, tuned to the strongly repulsive regime, known as impenetrable-bosons or Tonks-Girardeau (TG)  limit. This regime has been experimentally achieved \cite{Paredes04,Kinoshita04,Kinoshita06,Palzer09,VanDruten08,Bouchoule2011}. From the theoretical point of view, a mapping onto a Fermi gas \cite{Gir1960} yields the exact  many-body wavefunction of the system, allowing to obtain with extreme accuracy information on the state of the system and on its quantum dynamics. In fact, it is its {\em fermionic} character which ensures the  stability of the system in the experiment \cite{GanShlyap03}.
In this work we explore some of the universal aspects of the 1D Bose gas in the TG regime. In particular, we focus on the Tan's contact coefficient \cite{Tan2008c}, which corresponds to a  two-body correlation function. Being associated to the average interaction energy in the system, it directly reflects the universal aspects of the system once the interactions are tuned to very large values. The Tan's contact can be extracted from the wings of the momentum distribution \cite{Tan2008c,Zwe11}. Momentum distributions have been accurately measured in the  experiment \cite{Jacqmin2012}. 

A very important experimental issue is the effect of the temperature. The various finite-temperature regimes of the 1D Bose gas have been previously identified  \cite{Kheruntsyan2003,BouKerShl07,Jac11}. However, the momentum distribution of a 1D Bose gas at finite temperature has received relatively little attention till now. For a homogeneous Bose gas it is in principle accessible by thermal Bethe Ansatz calculations by extension of Ref.~\cite{Kozlowski2011}.  From the large-distance properties of the one-body density matrix, the Luttinger-liquid model predicts a Lorentzian shape at small momenta
\cite{Giamarchi_book}. 
The momentum distribution for a TG  gas was studied in a seminal work by Lenard \cite{Len66}, who was mainly interested in the thermodynamic limit.
The momentum distribution of a TG gas  under both box trap and harmonic confinement has been  numerically evaluated through  lattice simulations for hard-core bosons at low filling \cite{Rig05}. We obtain here for the first time an expression for  the tails of the  momentum distribution at finite temperature  and  extract the universal contact coefficient. Our approach is valid for any external confinement, and is  mostly analytical. As a main result, we obtain that the weight of the universal tails of the momentum distribution {\em increases} with temperature.

{\it One-body density matrix from the thermal Bose-Fermi mapping.} 
We consider  $N_B$ bosons of mass $m$
 confined by the harmonic  potential $V(x)=m\omega^2 x^2/2$. The
particles interact via the contact potential $v(x)=g\delta(x)$, and the Hamiltonian is given by 
\begin{equation}
{\cal H}=\sum_j\left[ -\frac{\hbar^2}{2m} \frac{\partial^2}{\partial x_j^2} + V(x_j)\right]+ g \sum_{j<\ell} \delta(x_j-x_\ell)
\end{equation}
In this work we focus on the impenetrable limit $g\to
\infty$, where  the effect of contact interactions can be  replaced by a cusp condition on 
the many-body wavefunction. The repulsions are so strong that the many-body wavefunction  vanishes when two particles meet, ie $\Psi(...x_j=x_\ell...)=0$ for each pair $\{j,\ell\}$.   The exact  solution of the many-body Schroedinger equation ${\cal H}\Psi_{N,\alpha}=E_{N,\alpha}\Psi_{N,\alpha} $ for any $N$-particle state individuated by the quantum numbers $\alpha=\{\nu_1,...\nu_N\}$ satisfying the above cusp condition is readily obtained by the  Bose-Fermi mapping \cite{GirDas02}. This  allows to obtain the bosonic wavefunction in terms of the one of an ideal Fermi gas in the same external potential and with the same quantum numbers, times a mapping function which ensures bosonic symmetry under particle exchange,
\begin{equation}
\label{Eq:psi_TG}
\Psi_{N,\alpha}(x_1...x_N)=\Pi_{1\le j<\ell\le N} {\rm sign}(x_j-x_\ell) \Psi^F_{N, \alpha}(x_1,x_2..,x_N).
\end{equation}
Here,  $  \Psi^F_{N,\alpha}(x_1,x_2..,x_N) = \frac{1}{\sqrt{N!}} \det[u_{\nu_j}(x_k)]$  is the fermionic wavefunction constructed with the  single particle orbitals $u_{\nu_j}(x)$. For harmonic oscillator confinement,  rescaling the spatial coordinate in units of the harmonic oscillator length   $a_{ho}=\sqrt{\hbar/m\omega}$, the single particle orbitals read  $u_{\nu}(x)= H_\nu(x) e^{-x^2/2}/\sqrt{2^\nu \nu!}/\pi^{1/4}$ with corresponding single-particle energies $\varepsilon_{\nu_j}=\hbar \omega (\nu_j+1/2)$. 

The above Bose-Fermi mapping allows to construct the thermal average of any observable. Observables that do not depend on the sign of the many-body wavefunctions are readily given by their fermionic counterparts \cite{GirDas02}, while substantial differences are expected for those observables which do depend on the sign of $\Psi$ as it is the case of the   
one-body density matrix. Its expression  at temperature $T$ in the grand-canonical ensemble reads
\begin{eqnarray}
\label{Eq:rho1_initial}
\rho_{1B}(x,y)&=&\sum_{N,\alpha} P_{N,\alpha} N \int_I {\rm d}x_2,...{\rm d}x_N 
\nonumber \\
&\times&
\Psi_{N, \alpha}(x,x_2..,x_N)\Psi_{N, \alpha}^*(y,x_2,..,x_N).
\end{eqnarray}
Here  $I=(-\infty, \infty)$ is  the spatial integration domain for harmonic confinement, $P_{N,\alpha}=e^{-\beta(E_{N,\alpha}-\mu N)}/Z$ is the thermal distribution function, $Z=\sum_{N,\alpha}e^{-\beta(E_{N,\alpha}-\mu N)}$ the partition function for the Tonks-Girardeau gas (or the mapped Fermi gas)  with  $E_{N,\alpha}=\sum_{j=1}^N \varepsilon_{\nu_j}$, $\beta=1/k_BT$, and $\mu$ the chemical potential.

The evaluation of Eq.(\ref{Eq:rho1_initial}) as it stands appears as a formidable task, but we simplify it as in an early work by Lenard  \cite{Len66} (see the supplementary material for details). The final result reads
\begin{eqnarray}
\label{lenard_series_final}
\rho_{1B}(x,y)&=&\sum_{j=0}^\infty \frac{(-2)^j}{j!} ({\rm sign}(x-y))^j \int_{x}^{y} {\rm d}x_2... {\rm d}x_{j+1} \nonumber \\
&\times& \det [\rho_{1F}(x_i,x_\ell)]_{i,\ell=1,j+1},
\end{eqnarray}
where $\rho_{1F}(x,y)=\sum_{j=1}^N f_{\nu_j} u_{\nu_j}(x)u_{\nu_j}^*(y)$ is the {\em fermionic} one-body density matrix,  $f_\nu=1/[e^{\beta(\varepsilon_\nu-\mu)}+1]$ is the Fermi occupation factor of a single-particle energy level, and 
in the above determinant one has to take $x_i=x$ for $i=1$ and $x_\ell=y$ for $\ell=1$. This illustrates how also for off-diagonal coherences of the Tonks-Girardeau gas it is possible to resort to the solution of a fermionic problem, thus  more involved than the one needed for the density profile.

{\em Momentum distribution of a Tonks-Girardeau gas under harmonic confinement.} 
Equation (\ref{lenard_series_final}) is our starting point for the evaluation of the bosonic one-body density matrix, which, by Fourier transform, yields the momentum distribution of the gas according to $n(k)=(1/2\pi)\int {\rm d}x\, 
{\rm d}y\, e^{i k (x-y)} \rho_{1B}(x,y)$. 

First of all, we proceed to further simplifying Eq.~(\ref{lenard_series_final}). Expanding the determinant in (\ref{lenard_series_final}) and using the definition of the fermionic one-body density matrix,  the $j$-th term of the one-body density matrix  according to the expansion $\rho_{1B}(x,y)= \sum_j [-2 \, {\rm sign}(x-y)]^j\rho_{1B}^{(j)}(x,y) /j!$   is given by
\begin{eqnarray}
\rho_{1B}^{(j)}(x,y)&=&\!\!\!\!\sum_{\nu_1..\nu_{j+1}}  f_{\nu_1}... f_{\nu_{j+1}}\!\!\! \sum_{P\in {\cal S}_{j+1}} (-1)^P u_{\nu_1}(x) u_{\nu_{P(1)}}(y)\nonumber \\
&& \prod_{\ell=2}^{j+1}\int_y^x \!\! \!\! \!\!dx_\ell \, u_{\nu_\ell}(x_\ell)  u_{\nu_{P(\ell)}}^*(x_\ell).
\end{eqnarray}
This can be finally casted onto the compact form 
\begin{equation}
\label{Eq:rho1_operative}
\rho_{1B}^{(j)}(x,y)=\!\!\!\!\sum_{\nu_1..\nu_{j+1}} \!\!\!\! f_{\nu_1}... f_{\nu_{j+1}} \sum_{k=1}^{j+1}u_{\nu_1}(x) A_{\nu_1\nu_k}(x,y) u_{\nu_k}^*(y)
\end{equation}
where
\begin{equation}
 A_{\nu_1\nu_k}(x,y)=(-1)^{1+k} \det {\bf B}_{\nu_1 \nu_k}(x,y)
\end{equation}
and 
$\det {\bf B}_{\nu_1 \nu_k}(x,y)$ is the minor determinant of the matrix function ${\bf B}(x,y)$  taking out the row 1 and column $k$, whose full expression is given in the supplementary material. We have thus considerably simplified the complexity of the problem, by reducing a $j$-variable multi-dimensional integral to the combination of  single-variable integrals, which moreover can be performed analytically in terms of special functions (see the supplementary material for details).  We are finally left to numerically evaluate only the sums over the single-particle levels in Eq.(\ref{Eq:rho1_operative}).

\begin{figure}
\centering
\includegraphics[width=1.0\linewidth]{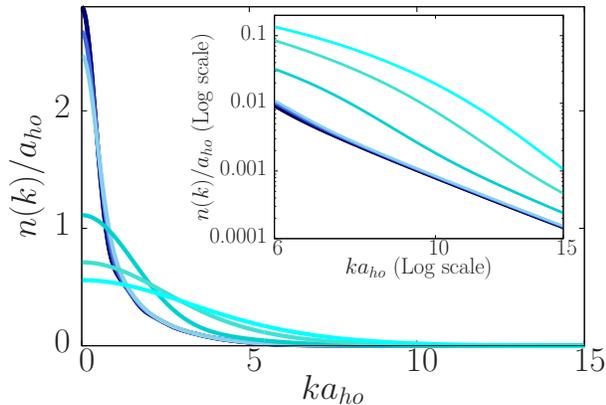}
\caption{\label{fig1} (Color online) Momentum distribution of a Tonks-Girardeau gas (in units $a_{ho}$) as a function of wavevector (in units of $a_{ho}^{-1}$) with $N=5$ particles  under harmonic confinement at increasing temperature, from top to bottom in the main peak $k_BT/\hbar \omega=0.1,0.5,0.7, 5,10,15$. The last three curves at high temperature are obtained by the use of the classical fermionic one-body density matrix (\ref{Eq:rho1_class}) into the series expansion (\ref{lenard_series_final}) up to $j=1$. The inset shows the same curves in double-logarithmic scale.}
\end{figure}

The momentum distribution of the TG gas at finite temperature is illustrated in Fig.\ref{fig1} \footnote{To keep a short computational time, the calculations are performed for $N=5$. Ref.\cite{Rig05} shows that even at small $N$  the grand-canonical description is reasonably accurate.}. At increasing temperature we notice a decrease of the peak at small momenta (this was also observed in the numerical simulations \cite{Rig05}), while at small temperatures the wings of the distributions appear to be less affected by the thermal excitations. The high-momentum tails  of the distribution are highlighted in the inset, where the momentum distribution is shown in double-logarithmic scale. As a main result, we find that the weight of the tails increases at increasing temperature. This is remarkably different from the case of a unitary three-dimensional Fermi gas. As we shall see below, this difference stems from the effect of reduced dimensionality.

{\em High-momentum tails at finite temperature} 
The momentum distribution of a 1D Bose with contact interactions displays at high momenta a universal power-law decay \cite{Minguzzi02,OlsDun03}  $n(k)\to {\cal C}/k^4$ for $k\to \infty$, independent on interaction strength. The weight ${\cal C}$ is now known as the Tan's contact \cite{Tan2008c}, and is related to quantum average of the interaction energy ${\cal H}_{int}= g \sum_{j<\ell} \delta(x_j-x_\ell)$,
\begin{equation}
{\cal C}=\frac{g m^2}{\hbar^4} \langle {\cal H}_{int}\rangle
\end{equation}
which is a  two-body correlator \cite{OlsDun03}. This remains finite in the TG limit $g\to \infty$ due to the simultaneous vanishing of the zero-distance density-density correlations and of the 1D scattering length. Its expression in the TG limit can be found  in \cite{Fang09}.
   
We start again from Eq.(\ref{lenard_series_final}) to estimate the weight of the high-momentum tails. They are related to the short-distance non-analytic behaviour of the one-body density matrix induced by the presence of the delta-interactions. The task is greatly simplified by the fact that the only term contributing to the high-momentum tails in the series for $\rho_{1B}(x,y)$ is the $j=1$ term, which upon short-distance expansion yields  \footnote{This follows immediately by noticing that the $j=0$ (fermionic) term is nonsingular at short distances and that all terms $j\ge 2$ contribute to higher order in the short-distance expansion} 
\begin{equation}
\label{Eq:FDef}
\rho_{1B}^{(j=1)}(x,y)\sim \frac{|x-y|^3}{3} F(R),
\end{equation}
where we introduced the center-of-mass coordinate  $R=(x+y)/2$ and the two-body function $F(R)$ is given by  
\begin{equation}
\label{Eq:FHermite}
F(R)= n(R) \sum_\nu f_\nu |\partial_R u_\nu(R)|^2-|\sum_\nu f_\nu u_\nu(R)\partial_R u_\nu^*(R)|^2
\end{equation}
with $n(R)=\sum_\nu f_\nu |u_\nu(R)|^2$ being the density profile.
Using the property of the asymptotics of  Fourier transforms, $\int {\rm d}z e^{-ik(z-z_0)} |z-z_0|^{\alpha-1} F(z) \rightarrow \frac{2}{|k|^\alpha}F(z_0) \cos(\pi\alpha/2) \Gamma(\alpha)$ for $k\to \infty$ we finally obtain  the expression for the contact 
\begin{equation}
\label{Eq:C_operative}
{\cal C}=\frac{2}{\pi}\int {\rm d}R \, F(R).
\end{equation}

At high temperatures, to estimate   the high-momentum tails 
we  use the classical limit for $\rho_{1F}(x,y)$ 
\begin{eqnarray}
\label{Eq:rho1_class}
\rho_{1F}(x,y)\simeq {\cal A}\exp \left[-\frac{(x+y)^2}{4} {\rm tanh} \frac{\beta \hbar \omega}{2} \right. \nonumber \\ \left.- \frac{(x-y)^2}{4} {\rm coth} \frac{\beta \hbar \omega}{2}\right],
\end{eqnarray}
with ${\cal A}=e^{\beta \mu} e^{-\beta \hbar \omega/2}/[\pi(1-e^{-2\beta\hbar\omega})]^{1/2}$, to construct 
the $j=1$ term in the sum (\ref{lenard_series_final}) for $\rho_{1B}(x,y)$.
From its short-distance expansion, the classical limit of the two-body function reads
 \begin{equation}
\label{Eq:F_class}
F^{class}(R)\simeq \frac{{\cal A}^2}{2} {\rm coth}\frac{\beta \hbar \omega}{2} \exp\left(-2 R^2  {\rm tanh} \frac{\beta \hbar \omega}{2} \right).
\end{equation}
The large-temperature behaviour of the contact is then obtained from Eq.(\ref{Eq:C_operative})  and (\ref{Eq:F_class}) as
\begin{equation}
{\cal C}^{class}=\frac{N^2}{\pi^{3/2}}\sqrt{\frac{k_BT}{\hbar \omega}}.
\label{Eq:contact_class}
\end{equation}
We thus find that a classical Tonks-Girardeau gas in harmonic confinement displays large-momentum tails   with a weight which increases as the square-root of the temperature. On the other hand, at increasing temperature the wavevector  $k_0$ starting from which the momentum distribution displays algebraic tails also increases with temperature, ie  $k_0\propto 1/\lambda_{dB}$ in the classical regime, with  $\lambda_{dB}=\sqrt{2 \pi \hbar^2/mk_BT}$. We would also like to notice that the behaviour of the contact at large temperatures depends on the type of external  confinement; for a box trap a similar calculation yields that the contact increases linearly with temperature. 

\begin{figure}
\centering
\includegraphics[width=1.0\linewidth]{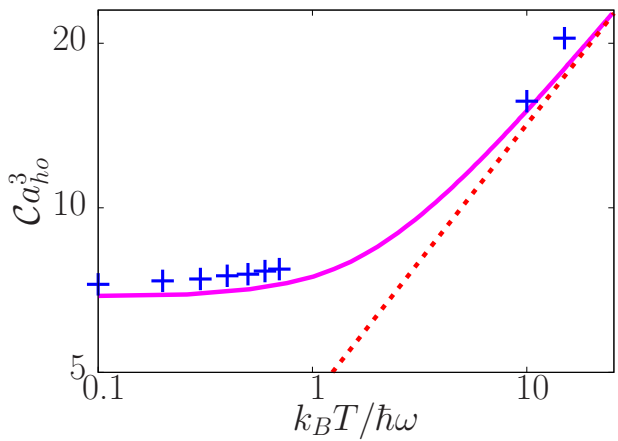}
\caption{\label{fig2} (Color online) Tan's contact (in units of $a_{ho}^3$) as a function of temperature $k_BT$ (in units of $\hbar \omega$)  for a TG gas under harmonic confinement. The full expression from (\ref{Eq:FHermite}) and (\ref{Eq:C_operative}) (solid, magenta) is compared with the high-temperature limit  (\ref{Eq:contact_class}) (dashed, red) and the data from the  numerical calculation of the momentum distribution (crosses, blue). 
}
\end{figure}

The temperature dependence of the  contact obtained from Eqs.(\ref{Eq:FHermite}) and (\ref{Eq:C_operative}) is illustrated in Fig.\ref{fig2}, and compared to the results from numerical evaluation of the full momentum distribution and the high-temperature limit~(\ref{Eq:contact_class}).  

{\em Virial approach for the contact at high temperature} 
To gain further insight in the large-temperature behaviour of the contact, we derive it using a virial approach. 

The finite-temperature contact can  be obtained  from  from Tan's sweep relation \cite{Tan2008c,Zwe11} in its thermodynamic form \cite{Hui11}
\begin{equation}
\label{Eq:sweep_thermo}
\left.\frac{d \Omega}{d a_{1D}}\right|_{\mu,T}=\frac{\hbar^2}{2m} {\cal C}
\end{equation}
where 
$a_{1D}=-2 \hbar^2/m g$ and $\Omega$ is the grandthermodynamic potential. Eq.(\ref{Eq:sweep_thermo}) follows from the Hellmann-Feynman relation together with the thermodynamic identity relating energy and grandthermodynamic potential  \cite{Hui11}. The advantage of this formulation is that it yields  a virial expansion form for the contact.

We start from the virial expansion of the grandthermodynamic potential $\Omega=- k_BT Q_1 (z +b_2 z^2 + b_3 z^3 +....)$ in powers of the fugacity  $z=e^{\beta \mu}$,  with  $b_2=\frac{Q_2}{Q_1}-\frac{Q_1}{2}$, $Q_2=Q_1 \sum_\nu e^{-\beta \epsilon_\nu^{\rm rel}}$  and more generally $Q_n={\rm Tr} e^{-\beta{\cal H}_n}$, is the $n$-body cluster. The energies  $ \epsilon_\nu^{\rm rel}$ entering in $b_2$ 
 correspond to the eigenvalues of the two-body problem in the relative-coordinate variable. Using the Tan's sweep relation, the above virial expansion turns then into the virial expansion for the contact, 
\begin{equation}
{\cal C}=\frac{2m}{\hbar^2 \lambda_{dB}} k_BT Q_1 (c_2 z^2 + c_3 z^3 + ...)
\end{equation}
with adimensional coefficients given by 
\begin{equation}
\label{Eq:cn}
c_n=-\frac{\partial b_n}{\partial (a_{1D}/\lambda_{dB})}.
\end{equation}
As a consequence of the scale invariance associated to the Tonks-Girardeau limit we can predict a  universal feature of the coefficients   $c_n$, namely we expect that they are constant (ie independent on temperature) since  $ (a_{1D}/\lambda_{dB})$ vanishes in the TG limit.

The direct evaluation of $c_2$  in the TG limit follows. In harmonic trap we have that $\epsilon_\nu^{\rm rel}=\hbar \omega(\nu+1/2) $ with $\nu$ given by the solution of the transcendental equation \cite{Busch1998}
\begin{equation}
\frac{\Gamma(-\nu/2)}{\Gamma(-\nu/2+1/2)}=\frac{\sqrt{2}a_{1D}}{a_{ho}}.
\end{equation}
The TG regime corresponds to $a_{1D}=0$, hence  $\nu=2n+1$ with $n=0,1,2,3...$ and the derivative required in (\ref{Eq:cn}) is readily evaluated. We finally obtain 
\begin{equation}
c_2=\frac{2^{3/2} \beta \hbar \omega  \lambda_{dB}}{\pi a_{ho}} \sum_{n=0}^\infty \frac{\Gamma(n+3/2)}{n!} e^{-\beta \hbar \omega (2 n +3/2)}.
\end{equation}
The sum is performed exactly to obtain
\begin{equation}
c_2=\frac{2 (\beta \hbar \omega)^{3/2}e^{-2\beta \hbar \omega}}{[e^{\beta \hbar \omega}-e^{-\beta \hbar \omega}]^{3/2}}  \to \frac{1}{\sqrt{2}} \,\,\,\, {\rm for}  \,\,\,\, {\beta \hbar \omega \ll 1}.
\end{equation}
We hence find a constant  value of the two-body contact coefficient at large temperature, as expected from universality considerations.  Using the above result for $c_2$, and the classical expression for $Q_1=k_BT/\hbar \omega$ and for the fugacity $z=N \hbar \Omega/k_BT$, we readily recover  Eq.(\ref{Eq:contact_class}).

{\em Conclusions.} We have developed the formalism of the thermal Bose-Fermi mapping to calculate  the momentum distribution of a Tonks-Girardeau gas under harmonic confinement at finite temperature. As a main feature, we have found that its high-momentum tails have a weight -- the Tan's contact -- which increases with temperature, and have linked this to the effect of reduced dimensionality. Moreover, we have shown that this result can be understood as one of the signatures of universality in TG gases, associated to scale invariance in the infinite interaction limit. Our results seem within experimental reach \cite{Haller2011,Jacqmin2012} and open the way to exploring further aspects of universality in strongly interacting 1D gases, both homogeneous and under confinement.

\acknowledgments We acknowledge discussions with I.~Bouchoule, B.~Fang and J.-S. Caux. AM acknowledges support from the Handy-Q ERC project.

\begin{widetext}

\section{{\bf Supplementary Material for: Universal contact for a Tonks-Girardeau gas at finite temperature}}

{\em \bf Derivation of the fermionic expression for the {\em bosonic} one-body density matrix}

We derive here the expression (4) for the bosonic one-body density matrix at finite temperature in terms of fermionic quantities. As a first step, due to the effect of the mapping function in Eq.(2) of the main text
each  many-body integral in Eq.(3) of the main text 
has the form
\begin{eqnarray}
\int_I dx_2,...dx_N \Psi_{N,\alpha}(x,x_2..,x_N)\Psi_{N,\alpha}^*(y,x_2,..,x_N) =\prod_{i=2}^N \int_I {\rm d}x_i {\rm sign}(x-x_i)\, {\rm sign}(y-x_i) f(x,y,x_2,...x_N)
\end{eqnarray}
where for short-hand notation we have set $f(x,y,x_2...x_N) = \Psi^F_{N,\alpha}(x,x_2...x_N) \Psi_{N,\alpha}^{F*}(y,x_2...x_N)$. 
We use that for each integration variable $ \int_I {\rm d}x_i {\rm sign}(x-x_i)\, {\rm sign}(y-x_i) \, f= \int_I dx_i \,  f -2 \,{\rm sign}(x-y) \int_y^x dx_i\, f $ and the binomial power series to rewrite Eq.(3) of the main text as
\begin{eqnarray}
\label{Eq:lenard_interm}
\rho_{1B}(x,y)&=&  \sum_{N,\alpha} P_{N,\alpha}\, N\sum_{j=0}^{N-1} \binom{N-1}{j} 
(-2)^j ({\rm sign}(x-y))^j \int_{x}^{y} {\rm d}x_2... {\rm d}x_{j+1} 
\nonumber \\  \nonumber
&\times& 
  \int_I dx_{j+2}... dx_N\Psi^F_{N, \alpha}(x,x_2..,x_N)\Psi_{N,\alpha}^{F*}(y,x_2,..,x_N).
\end{eqnarray}
Here, we recognize the fermionic $j$-body density matrices, which are defined  as
\begin{eqnarray}
\rho_{jF}(x_1..x_j;x_1'..x_j')=\sum_{N,\alpha} P_{N,\alpha} \frac{N!}{(N-j)!}  \int_I {\rm d}x_{j+1}... {\rm d}x_N \Psi^F_{N,\alpha}(x_1..x_{j+1}..x_N)\Psi_{N,\alpha}^{F*}(x_1'..,x_{j+1}..,x_N).
\end{eqnarray}
Thus, we are led to the  expression for the {\em bosonic} one-body density matrix as a sum of integrals of {\em fermionic} density matrices,
\begin{eqnarray}
\label{lenard_series}
\rho_{1B}(x,y)&=&\sum_{j=0}^\infty \frac{(-2)^j}{j!} ({\rm sign}(x-y))^j \int_{x}^{y} {\rm d}x_2... {\rm d}x_{j+1}   
\rho_{j+1,F}(x,x_2,...x_{j+1};y,x_2,...x_{j+1}).
\end{eqnarray}
Further progress can be made by noticing that for noninteracting fermions the $j$-body density matrices factorize in terms of the corresponding {\em one-body} density matrices according to
\begin{equation}
\rho_{jF}(x_1,x_2,...x_n;x_1',x_2',...x_n')=\det [\rho_{1F}(x_i,x_\ell')]_{i,\ell=1,n},
\end{equation}
where $\rho_{1F}(x,y)=\sum_{j=1}^N f_{\nu_j} u_{\nu_j}(x)u_{\nu_j}^*(y)$ and $f_\nu=1/[e^{\beta(\varepsilon_\nu-\mu)}+1]$. The final expression for the bosonic one-body density matrix (4) in the main text
 is thus obtained.

\vspace{0.3cm}
{\em \bf Matrix functions $B_{\nu_i\nu_j}$ entering the determinantal expression of the one-body density matrix under harmonic confinement}

In the main text we have demonstrated how the initial multi-variable integration needed to compute the one-body density matrix  as given by Eq. (4) of the main text  can be reduced to the calculation of the minor determinant of the matrix ${\bf B}(x,y)$. Here we give its full expression for the case of a TG gas under external harmonic confinement.

As it is readily obtained from Eqs.(5) and (6)  of the main text, the expression for the matrix ${\bf B}(x,y)$ is given by
\begin{equation}  
B_{\nu_i\nu_j}(x,y)={\rm sign}(x-y)\int_y^x dw \, u_{\nu_i}(w)u_{\nu_j}^*(w).
\end{equation}
For harmonic confinement, the calculation of the one-body density matrix can be further simplified by the fact that the elements of the matrix $B_{\nu_i\nu_j}(x,y)$ can be evaluated analytically.
Using the power series expansion of the Hermite polynomials $H_\nu(x)=\nu! \sum_{k=0}^{[\nu/2]}(-1)^k (2 x)^{\nu-2k}/k!/(\nu-2k)!$ and the result $\mu_m(x)=\int^x dw e^{-w^2} w^m= x^{m+1} e^{-x^2} _1F_1(1,(m+3)/2, x^2)$ where $_1F_1(a,b,c)$ is the confluent  hypergeometric function, we have
\begin{equation}
B_{\nu_i\nu_j}(x,y)=\frac{1}{\pi}\sqrt{\nu_i!\nu_j!2^{\nu_i+\nu_j}}\sum_{k}^{[\nu_i/2]}\sum_{k'}^{[\nu_j/2]}\frac{(-1)^{k+k'}}{k!k'! (\nu-2k)!(\nu-2k')!}\mu_{\nu_i+\nu_j-2(k+k')}(x,y).
\end{equation}
\end{widetext} 

\end{document}